# Data Sharing and Ontology use among Agricultural Genetics, Genomics, and Breeding Databases and Resources of the AgBioData Consortium

Jennifer L Clarke*[1], Laurel D. Cooper*[2], Monica F Poelchau*[3], Tanya Z. Berardini[4], Justin Elser[2], Andrew D Farmer[5], Stephen Ficklin[6], Sunita Kumari[7], Marie-Angélique Laporte[8], Rex T. Nelson[9], Rie Sadohara[10], Peter Selby[11], Anne E Thessen[12], Brandon Whitehead[13], Taner Z. Sen[#14]

1.  Department of Statistics, Department of Food Science and Technology, University of Nebraska-Lincoln, Lincoln, NE, 68583 USA
2.  Department of Botany and Plant Pathology, 2503 Cordley Hall, Oregon State University, Corvallis, OR, 97330, USA
3.  USDA, Agricultural Research Service, National Agricultural Library, 10301 Baltimore Ave, Beltsville, MD, 20705, USA
4.  TAIR and Phoenix Bioinformatics, USA
5.  National Center for Genome Resources, Santa Fe, NM 87505, USA
6.  Department of Horticulture, Washington State University, Pullman, WA 99164, USA
7.  Cold Spring Harbor Laboratory, Cold Spring Harbor, NY 11724, USA
8.  Digital Inclusion, Bioversity International, 34397 Montpellier, France
9.  USDA, Agricultural Research Service, Corn Insects and Crop Genetics Research Unit, Ames, IA, USA
10. Department of Plant, Soil, and Microbial Sciences, Michigan State University, 1066 Bogue St, East Lansing, MI, 48824, USA
11. Cornell University, Ithaca, NY 14850, USA
12. Department of Biomedical Informatics, University of Colorado Anschutz, 1890 N. Revere Court, Mailstop F600, Aurora CO, 80045 USA
13. Manaaki Whenua -- Landcare Research, Ltd., Palmerston North, New Zealand
14. USDA, Agricultural Research Service, Crop Improvement Genetics Research Unit, Western Regional Research Center, 800 Buchanan St, Albany, CA, 94710, USA and Department of Bioengineering, University of California, Berkeley, CA, 94720, USA

* joint first co-authors
# corresponding author

## KEYWORDS

# ABSTRACT

Over the last several decades, there has been rapid growth in the number and scope of agricultural genetics, genomics and breeding (GGB) databases and resources. The AgBioData Consortium (https://www.agbiodata.org/) currently represents 44 databases and resources covering model or crop plant and animal GGB data, ontologies, pathways, genetic variation and breeding platforms (referred to as 'databases' throughout). One of the goals of the Consortium is to facilitate FAIR (Findable, Accessible, Interoperable, and Reusable) data management and the integration of datasets which requires data sharing, along with structured vocabularies and/or ontologies. Two AgBioData working groups, focused on Data Sharing and Ontologies, conducted a survey to assess the status and future needs of the members in those areas. A total of 33 researchers responded to the survey, representing 37 databases. Results suggest that data sharing practices by AgBioData databases are in a healthy state, but it is not clear whether this is true for all metadata and data types across all databases; and that ontology use has not substantially changed since a similar survey was conducted in 2017. We recommend 1) providing training for database personnel in specific data sharing techniques, as well as in ontology use; 2) further study on what metadata is shared, and how well it is shared among databases; 3) promoting an understanding of data sharing and ontologies in the stakeholder community; 4) improving data sharing and ontologies for specific phenotypic data types and formats; and 5) lowering specific barriers to data sharing and ontology use, by identifying sustainability solutions, and the identification, promotion, or development of data standards. Combined, these improvements are likely to help AgBioData databases increase development efforts towards improved ontology use, and data sharing via programmatic means.

AgBioData databases: https://www.agbiodata.org/databases

# INTRODUCTION

The AgBioData Consortium (https://www.agbiodata.org/) consists of 44 agricultural genetics, genomics and breeding (GGB) databases and resources (https://www.agbiodata.org/databases; referred to as 'databases' throughout), along with their stakeholders and collaborators. The scope of the databases is quite broad, with a large representation from databases focusing on data from plants; including crops, trees and model species. Databases for domestic and livestock animals, fish, insects, microbes and pathogens are also represented, as well as resources for breeding tools, ontologies, pathways and genetic variation.

The goal of the Consortium is to increase collaboration and shared best practices among member databases. Founded in 2015, the Consortium was awarded a National Science Foundation Research Coordination Network (RCN) grant in 2021 (# 2126334) to strengthen and expand the network, in particular around the FAIR (Findable, Accessible, Interoperable and Reusable) principles (1). During the first year of the RCN grant, several working groups (WGs) were established around key priorities, and were tasked with defining actionable problems facing the community, and if possible, providing solutions.

The AgBioData Consortium GGB databases provide tools and services that allow scientists to discover, retrieve and reuse data, and are often curated and integrated by biocurators, with the goal of enhancing knowledge discovery and scientific progress (2,3). There are multiple aspects required for effective discovery, retrieval and reuse of data, including but not limited to the technology that exposes the data to end users or other systems, and annotations of the data to ontology terms that allow the data to be understood outside the context of the database, by both humans and computers. Optimization of both the technology used to expose the data, and ontologies used to understand the exposed data, is worthwhile, due to their central importance for GGB databases. The AgBioData Consortium formed two WGs focused on Data Sharing and Ontologies, respectively, in order to identify whether and how technology and ontology use in the Consortium could or should be improved. The Data Sharing and the Ontologies WGs concluded that an updated assessment of data sharing and/or ontology usage, problems, and needs in the AgBioData community was necessary in order to provide recommendations to the larger consortium on which problems should be prioritized.

**Data sharing** is a generic term covering all methodologies for passing information from one system to another, to be used by another person or tool (4). Data sharing methodologies exist across a spectrum rooted in how much automation is involved in the sharing process. At one end of the spectrum, there is manual sharing, such as shipping files to a collaborator on a hard drive. On the other end of the spectrum, there is full automation (federation) where data can travel between software systems freely and automatically, without human intervention. **Data federation** refers to a specific type of data sharing between data repositories that is highly automated and invisible to the end user (5). Data federation is driven by a shared data governance structure that is defined centrally, paired with distributed data hosting by local (or repository-specific) teams that have the autonomy to execute and enforce governance standards as appropriate for their specific repository. This allows for a high level of interoperability while ensuring security and compliance. In between these two ends of the spectrum, there are many other methods and technologies. Effective data sharing - among databases, and from databases to stakeholders - is a key component towards implementation of the FAIR principles (1). Key to effective data sharing is the ability to programmatically access data, such that data and metadata can be accessed by both humans and computers. The Data Federation/Sharing WG initially focused on data federation but expanded to include topics on data sharing.

Across the spectrum of agricultural databases and genomics resources, managers, curators and researchers are dealing with exponential increases in the volume of data to manage. One valuable class of tools that have emerged and grown over the past 20 to 25 years is that of ontologies. Starting in the late 1990's, the Gene Ontology (6,7) led the way in the biological sciences with the development of a resource which provided a structured, controlled vocabulary describing biological processes, molecular functions and cellular components. These ontology terms are related to one another through a controlled set of relations (8) and form a hierarchy which can be searched and reasoned across using computer software. In addition, ontology terms can be attached to data objects (annotation) in a standardized way, enabling integration and analysis across multiple studies or species (9).

Since the advent of the Gene Ontology (GO), there has been a massive growth in the number of biological ontologies, describing all facets of biology from anatomy (10–16) to chemicals (17), to traits and phenotypes (18–25) and environments (26,27), among many other domains.

In addition to the reference or species-neutral ontologies such as the GO, there has been a movement toward the development of species- or clade-specific controlled vocabularies (CVs) that are targeted towards breeding communities and their unique needs. A leader in this area, the Crop Ontology (CO; https://cropontology.org/) (28–30) is a collection CVs or trait dictionaries that cover plant anatomy, traits and phenotypes of approximately 35 plant species or clades (as of June 2023). The CO is widely adopted and has become an important component in the plant breeding databases such as the Integrated Breeding Platform (IBP; https://www.integratedbreeding.net/) Breeding Management System, and Breedbase (https://breedbase.org/) at the Boyce Thompson Institute (28). In addition, 11 of the CO trait dictionaries are currently mapped to the Plant Trait Ontology (TO) and integrated into the Planteome Database (https://planteome.org/) (18).

Sites such as the OBO Foundry (https://obofoundry.org/), the Planteome (https://planteome.org/), the NCBI Bioportal (https://bioportal.bioontology.org/) and the EMBL-EBI Ontology Lookup Service (https://www.ebi.ac.uk/ols/index) provide curated lists of ontologies for biological data curation and annotation. With the advent of the FAIR principles for data management (1) there is an increased need for the power of ontologies to facilitate data findability, accessibility, interoperability and repeatability.

# SURVEY GOALS AND METHODOLOGY

A previous survey, conducted in 2017 (31), assessed the status of data sharing and ontology use by the members of the AgBioData Consortium. Since five years had passed since the survey was delivered, the WGs saw an opportunity to gauge whether usage of ontologies or data sharing technologies had improved, and to identify new needs in the community that the RCN grant could address. The Data Sharing WG partnered with the Ontologies WG to construct and distribute a survey designed to gather information about current knowledge, use, and challenges to data federation and ontologies. The survey was created as a Google Form and shared with AgBioData members via email and the Slack application (https://slack.com/) in July 2022, and responses were collected through the end of August 2022 to assess the status and needs about data sharing and ontology use. Our target audience was database personnel, and our goal was to receive one response per member database (at the time, 44 databases). The survey can be viewed in Appendix 1. The survey was performed to learn the current practices and future plans of the AgBioData Consortium member databases, and as such, the findings should not be generalized to the larger universe of similar databases which are not AgBioData members. The Ontology results were compared to those found in the 2017 survey.

The main goals of the Data Sharing sections (6-10) of the survey were to assess:

1. the level of data sharing that AgBioData member databases *have* - what data and metadata are being shared; how they are being shared (covered in questions 6.1, 7.1, 7.2 and 7.3 of the survey)
2. the level of data sharing that AgBioData members *want* - is there a discrepancy between existing and desired data sharing for databases and their stakeholders (questions 9.1, 9.2, 9.3, 9.5, 9.6, 10.7, 10.8)
3. the barriers towards advancing to the desired data-sharing level (question 10.3)
4. AgBioData members' level of awareness of data sharing technologies, and need for or interest in training (question 8.1).

The Ontology sections (2-5) of the survey assessed the following:

1. the use of ontologies and/or CVs in member databases, both the publicly available resources and also in-house ones (questions 2.4, 3.1, 3.2)
2. the types of data which are annotated with the ontology or CV terms (question 3.3) and the tools in use to explore, browse or develop ontologies or CVs (question 3.4)
3. if the AgBioData members were exploring and/or contributing to ontology development and what tools were being used for that (questions 3.5, 3.6)
4. if the ontologies or CVs were integrated into member databases (question 3.7) and if they are used for search functions (question 3.8)
5. the reasons for not using ontologies and CVs in the member database (question 4.1), and what barriers to ontology utilization might exist (question 10.1)
6. the needs for training and education about ontologies (question 4.2)
7. the respondents' plans for ontology and/or CV use in the future (section 5)

# RESULTS AND DISCUSSION

## Survey Respondents and the Types of Data

We received 33 responses to the survey from individuals representing 37 GGB databases, as some respondents represented more than one database. All of the respondents are either members of the AgBioData Consortium, or represent a member database. Many of the databases host data across taxonomic kingdoms (e.g., a database can host both plant and animal data). The represented databases were skewed towards plant data. Crops, model plants and trees were represented by 25 databases. Eight databases provide data on livestock and/or domestic animals, and six on aquaculture and fisheries species. Six databases carried insect data, and another five data on vectors, microbes and parasites. Another four databases did not have a taxonomic focus.

Not all the respondents answered all the questions. It is important to note that in the survey results below, the percentage of respondents were not calculated based on the total number of respondents, but based on the number of respondents for that specific question.

The respondents serve a variety of roles within their databases/resources. Out of 33 respondents to this question, the survey takers identified themselves as project Principal Investigators (48.5%), curators (48.5%), maintainers (45.5%), developers (36.4%), or computational biologists (33.3%). Note that for this question, more than one answer was allowed.

We asked what data types each database handled. Over two-thirds of the 33 respondents reported working with genetic, genomic (sequence, transcriptome, markers, with annotations) and phenomic data, while several (4 or 12.1%) also reported working with epidemiological, nutrient, scientific text, or bibliography information. Of note was the diversity of the genetic and genomic data represented, from reference genome sequences and pangenomes to transcriptomics, QTLs, germplasm/breeding lines, mRNAs and proteins.

## Data Sharing Survey Results

### Current level of data sharing

Of the 32 respondents to this question, 87.5% reported that they currently share data with other databases, systems, or tools. The remaining 12.5% respondents stated that they either have the capability for sharing but it is not currently being used (9.4%) or it would take significant effort to enable sharing (3.1%). A caveat is that we did not follow up to ask how respondents that replied 'yes' know whether their data was being used. We conclude that AgBioData member and non-member databases do prioritize data sharing.

That said, 21.9% of respondents stated that sharing can only be done with specific tools. When asked about available mechanisms for data sharing, each of the following technologies were utilized by 60% of the 29 respondents: manual transfer of flat files (through FTP, email, or DropBox[TM]); hyperlinks to flat files; and discoverable web service APIs. A much smaller number of respondents use shared search indices (17%) or direct SQL access (10%), both of which require some expertise in a programming language and/or handling queries from database structures.

As can be seen in Figure 1A and B, most databases offer more than one well-recognized option for data sharing. Over 80% of those who offer manual transfer of flat files also provide a more automated option such as discoverable web service APIs. This serves as further evidence that AgBioData members do prioritize data sharing and federation. Overall, 25 out of 29 of respondents (86%) used technologies for data sharing where users could access the data programmatically (Discoverable Web Service APIs, Shared search indices, Direct SQL access (or alternate query language, e.g., GraphQL), iCommands - federated data storage and access, Ontology + RDF knowledge graph file, SPARQL endpoint, Specific web services). Four respondents did not answer the question. This suggests there is an opportunity to improve the extent of programmatic data access among AgBioData databases.

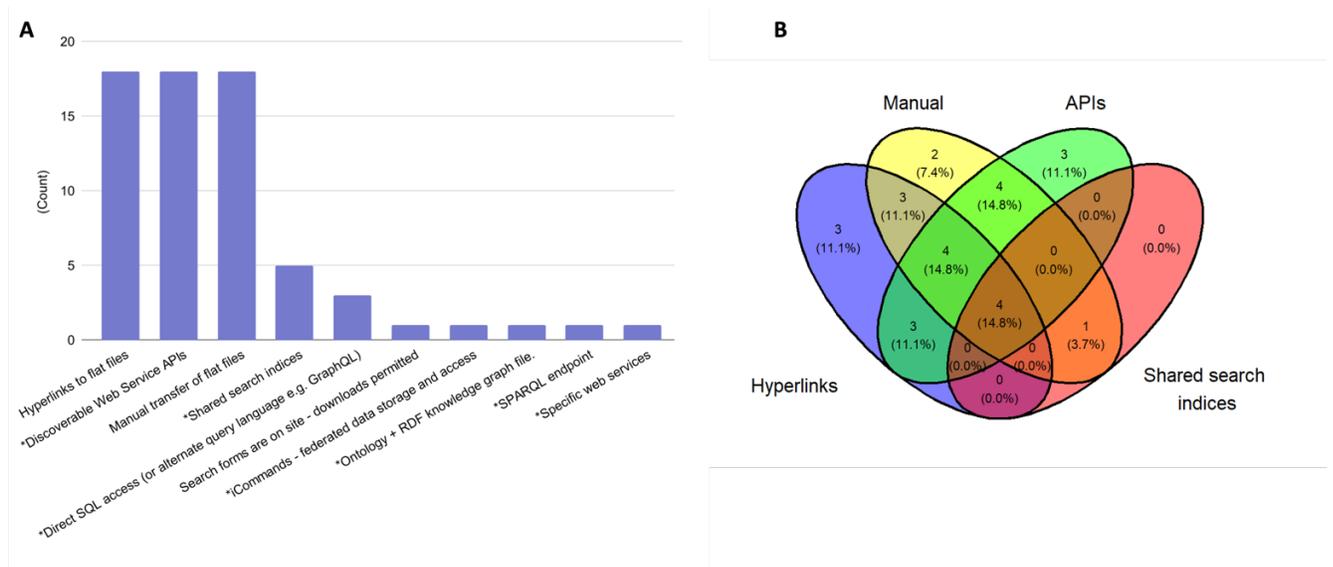

**Figure 1.** Mechanisms used for data sharing. There were 29 responses to this question. In Figure 1A, mechanisms that allow for programmatic access to the shared data are marked with asterisk (*). In Figure 1B, the four data types with the most responses are shown in a Venn diagram.

The survey also delved into the types of datasets shared with other databases (Fig. 3). The top three responses are: annotation/ontology files (69%), raw sequence files (52%), and variant files (41%). A third of databases reported sharing phenotypic data (e.g., images, traits) with another third reporting sharing marker data (e.g., PCR, SSR).

As we see in Figure 2A and B, of the most common genomic/genetic data types (annotation/ontology, raw sequence, variant, marker), almost 60% of databases represented in the responses received offer more than one data type with many offering 3 or more data types. Very few databases reported sharing ontologic data in the form of .obo or .owl files or homology information.

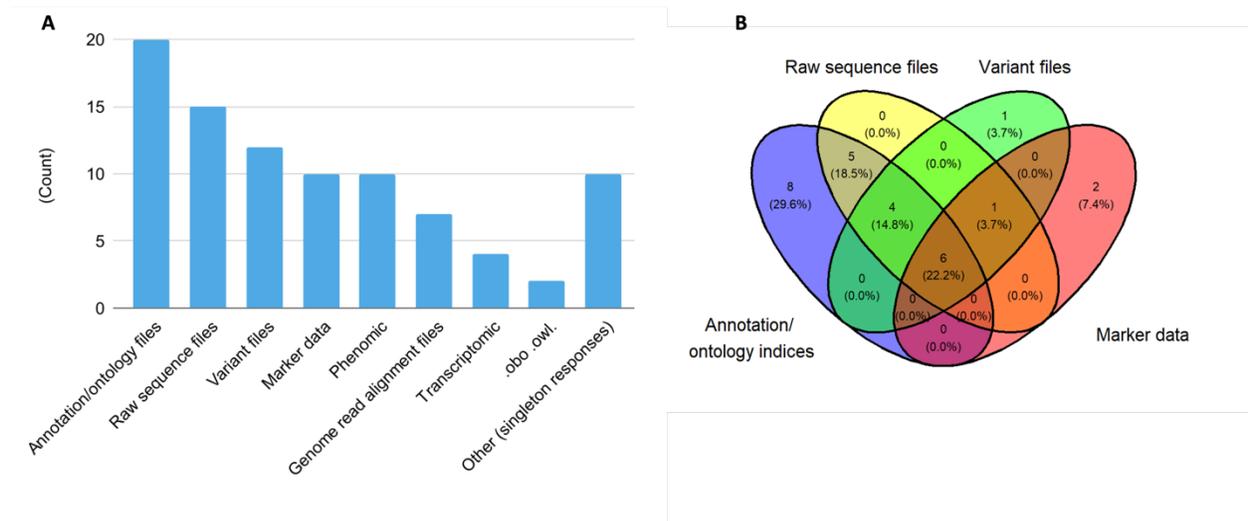

**Figure 2.** Data types shared with other databases. There were 29 responses to this question. In Figure 2A, singleton answers were collated into the 'Other' category. In Figure 2B, the four data types with the most responses are shown in a Venn diagram.

With respect to metadata, most databases reported the sharing of data-associated publications and/or the original data source (Fig. 3). Some databases (30%) do share information about the material and methods of data collection, and a few (10%) share environmental metadata. The survey shows that datasets are shared with information about associated publication(s) (73% out of 26 responses) and/or original data sources (69%). Hence, although sharing datasets is of primary importance, the metadata associated with these datasets are not always as informative. It is unclear whether metadata information is searchable, or if it is generally left to the user to uncover metadata from publications or original sources. While it may appear that over a quarter of respondents do not provide either associated publications or the original data source, all but two responses included at least one of the two. In some cases, more detailed information about materials and methods are provided (31%).

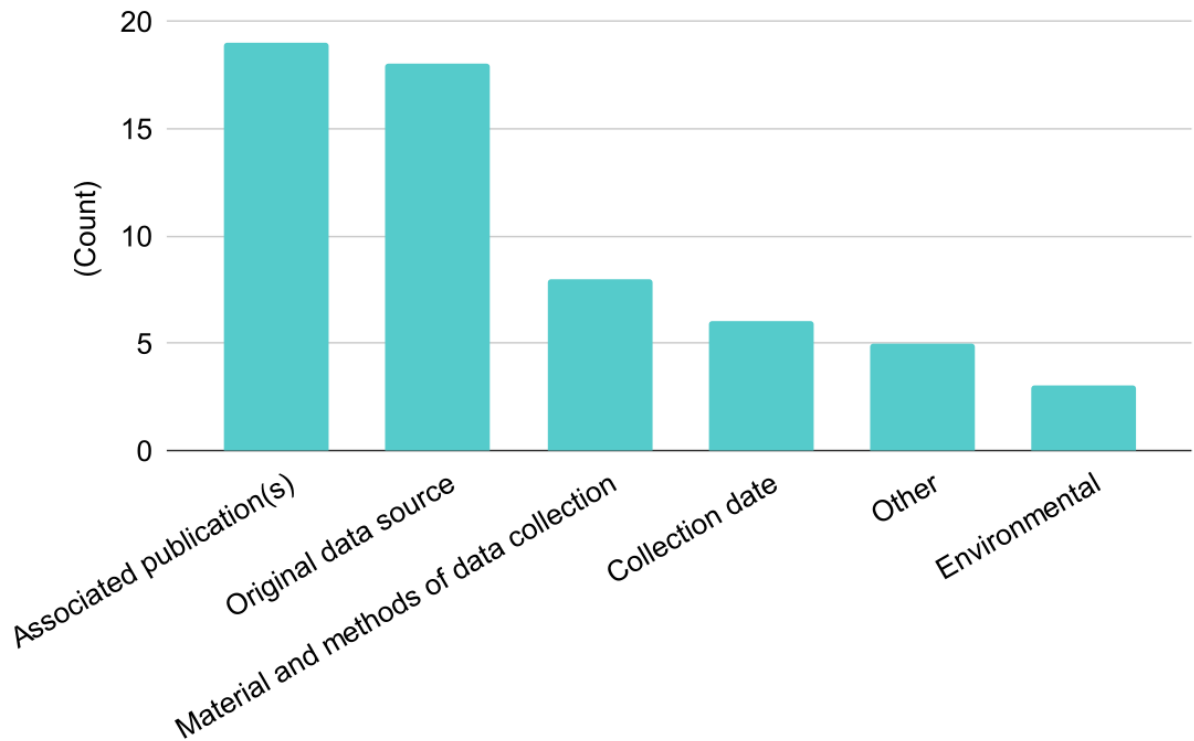

**Figure 3.** Types of metadata shared with other databases. There were 26 responses to this question. Singleton answers were collated into the 'Other' category, and metadata categories are abbreviated for this figure.

The responses reflect how ubiquitous and commonplace genome sequencing, annotation, and variant identification became for biological databases, along with their importance in research.

## Desired level of data sharing

Section 9 of the survey measured the willingness of biological databases to share their datasets. 91% of the 33 respondents put the importance of data sharing for their database either "very high or high." A follow-up question, however, showed that only 73% of the respondents think that their user community considers "very high/high" about the importance of data sharing (Fig. 4).

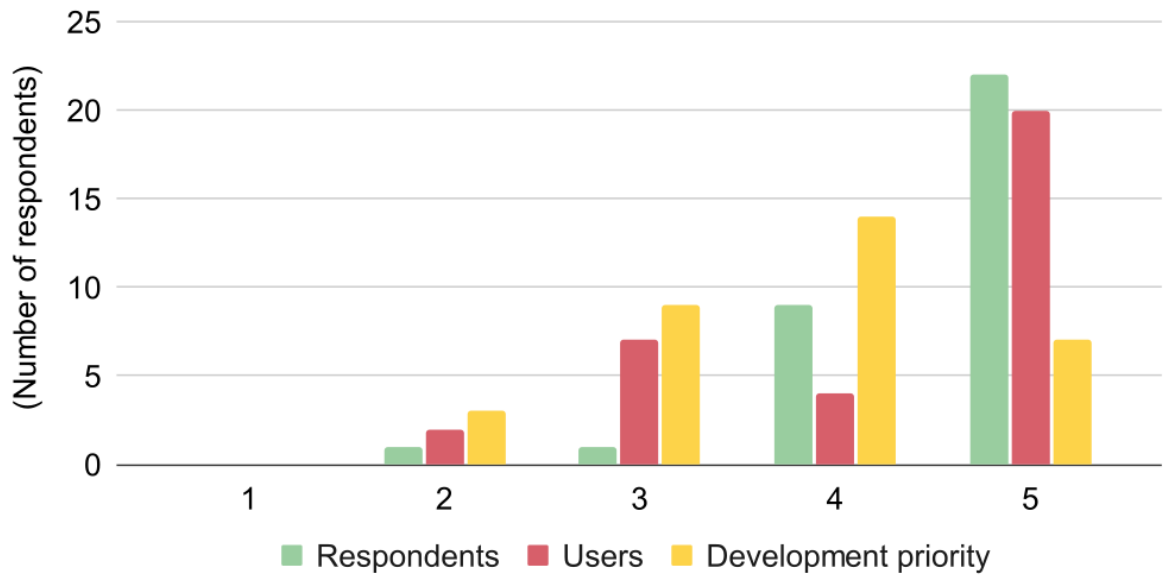

**Figure 4.** The importance of data discoverability and availability for respondents/databases (green), users (red), and development priorities (yellow). The respondents ranked the importance from 1 to 5, with 1 denoting the lowest importance and 5 denoting the highest importance.

However, when the respondents were reminded of the added costs of making data more shareable, the enthusiasm significantly shifts from very high/high to high/medium categories: data sharing was shown of very high importance by 67% of the respondents, but it is a high priority only for 21% of respondents. This is interesting as respondents were either open to suggestions about improving federation or had very concrete ideas regarding how they would like to make their databases more FAIR. These ideas included BrAPI compliant services, semantic web, GraphQL, and tailoring access for producers.

These seemingly contradictory results could be explained as follows: stakeholders are not pushing for sharing data across databases, and significant budget pressures restrict databases to only prioritize stakeholders' concerns. Hence, we attribute this drop in priority as a reflection of a lack of sustainable funding, i.e. the reality that most databases have fully allocated their available resources and therefore lack the capacity to make additional investments in development of any kind of data sharing. Perhaps increasing the awareness of the importance of data sharing across communities among stakeholders may shift the focus of databases towards devoting more funds for more data sharing.

We also asked what technologies or methodologies were desired to make data more available. Responses were free-form, and included using BrAPI compliant services; exploring GraphQL; microservices; and linked data/semantic web. Only one respondent was concerned with data privacy.

Questions 9.5, 9.6, 10.7 and 10.8 focused on types of data or metadata that the respondent would like to be able to share more effectively with external users, organizations, or tools; and conversely, that the respondent would like shared with them. As the response type was free-form, we classified the responses into categories, and tallied the number of responses per category. Responses from questions 9.5 and 10.7 were aggregated together, as were responses from questions 9.6 and 10.8, due to the similarity of the questions. The full results can be seen in Figure 5. There were 32 different data types given as answers. Among the most common responses were phenotypic or phenomic data (11 responses from 7 databases); all data types; gene data and metadata; and genomic data. The largest category of responses was 'singleton' answers (24 total). There were not many differences between data that respondents wanted to share from their own databases, vs. data they would like to access from others. The main distinction is that three respondents wanted to access GWAS data from other databases. Also, lack of consistent formatting, or applications of standards among databases, hampered data access from other databases. These results suggest two things: first, that the strongest need among AgBioData databases is for standards for phenotypic data sharing. Second, given the large number of singleton answers, that AgBioData databases provide access to a diverse set of data types, and that therefore we encounter diverse problems with data sharing.

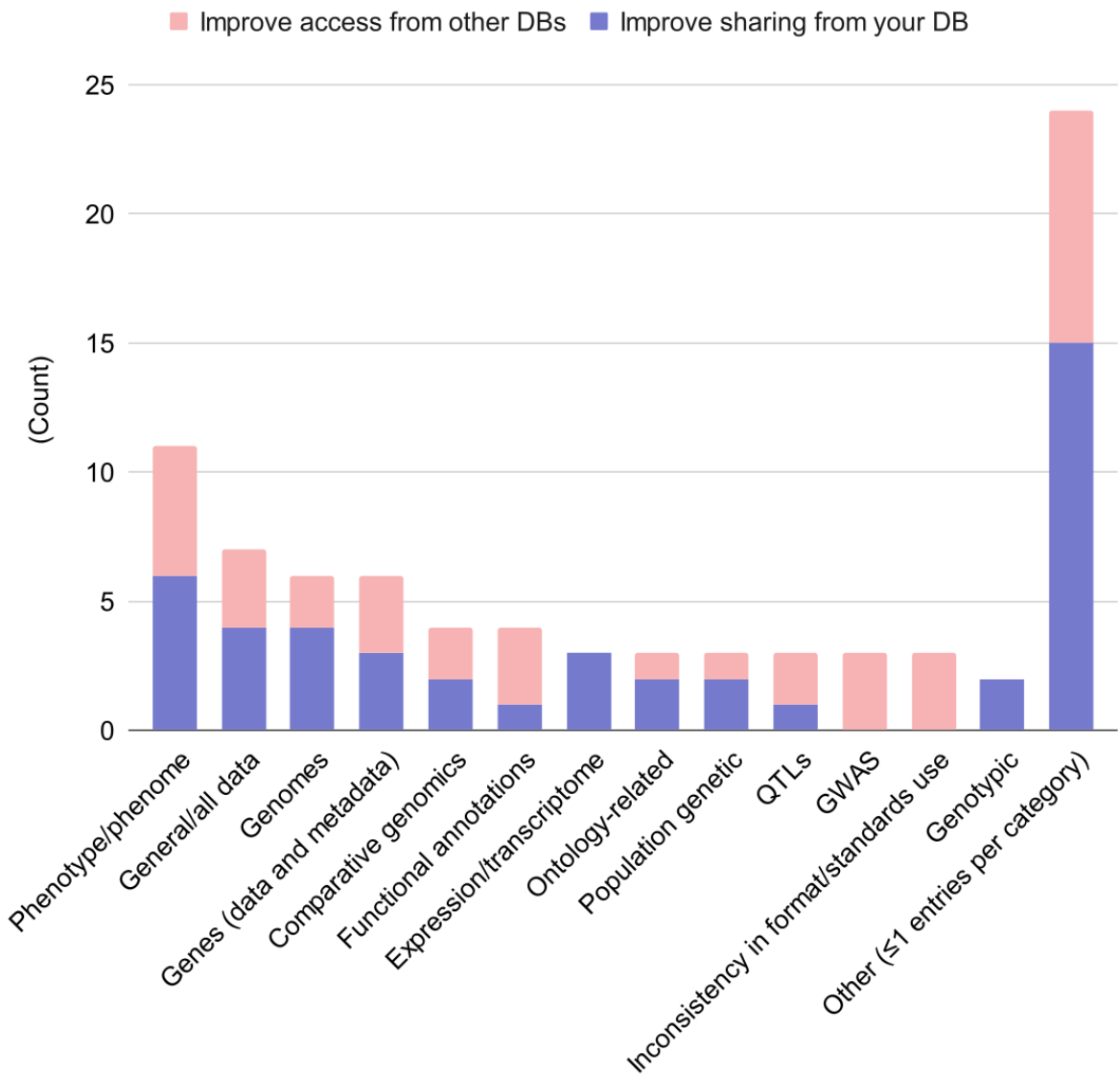

**Figure 5.** Desired data and metadata types shared among databases. Free-form answers from the questions 9.5 and 10.7 (what data and metadata would you like to share) were aggregated, with responses in blue; and free-form answers from questions 9.6 and 10.8 (what data and metadata would you like to access from other databases) were aggregated, with responses in red.

## Barriers to success

The 32 respondents identified the following as the top four barriers in data sharing: time and resources (81%), funding and return on investments (62%), lack of data standards (50%), and technical knowledge (47%). The rest of the barriers were all below 12% (Fig. 6). We recognize that technology can be used to reduce the time and resources needed to do the same amount of work - so these barriers are not necessarily independent categories and could be related. Further, especially for those who work in academia, it is not surprising to see funding/resources as the top barrier to get things done. Usually, increasing resources is out of the hands of researchers (one counter-example may be Phoenix Bioinformatics raising funds from its users to manage TAIR (32). However, the remaining top barriers, i.e., lack of data standards and technical knowledge, are areas that can be remedied through developing data standards and training, where AgBioData can play a significant role.

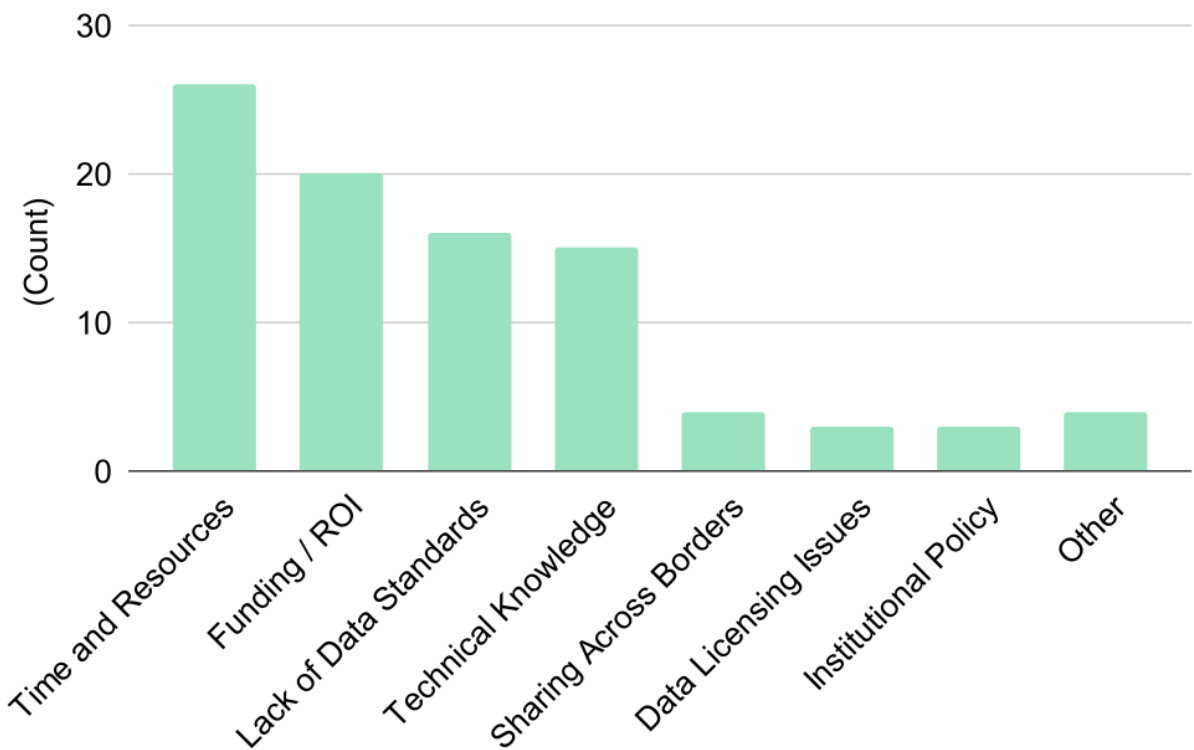

**Figure 6.** Blockers to successful data sharing. There were 32 responses to this question.

In spite of the barriers to data sharing mentioned above a majority of respondents - 87% (out of 23 responses) - have imported, linked, or shared data programmatically from another database. The respondents were mostly successful in their endeavors, which is encouraging. Most cautionary comments referred to instability of target URLs or data sources; and identifier ambiguity.

## Awareness of tools and technologies

All respondents reported a familiarity with manual file sharing, data hyperlinks and digital object identifiers (DOIs), and over half of respondents reported familiarity with APIs (either discoverable, RESTful, or programmatic) (Fig. 1A, 1B). However, two/thirds of respondents reported a lack of familiarity with other data federation tools such as linked data (e.g., semantic web), client integration of multiple results (e.g., sharing genome browser tracks), index driven search technologies (e.g., ElasticSearch), and large file transfer via services like Globus (Fig. 7). Fortunately, respondents reported a desire to learn more about these tools and technologies (such as GraphQL). This desire provides the data federation WG with an opportunity to promote and improve data federation through educational offerings. The WG also recommends that data sharing tools should be made intuitive enough to be used by non-programmers, instead of designed only for people with heavy programming experience.

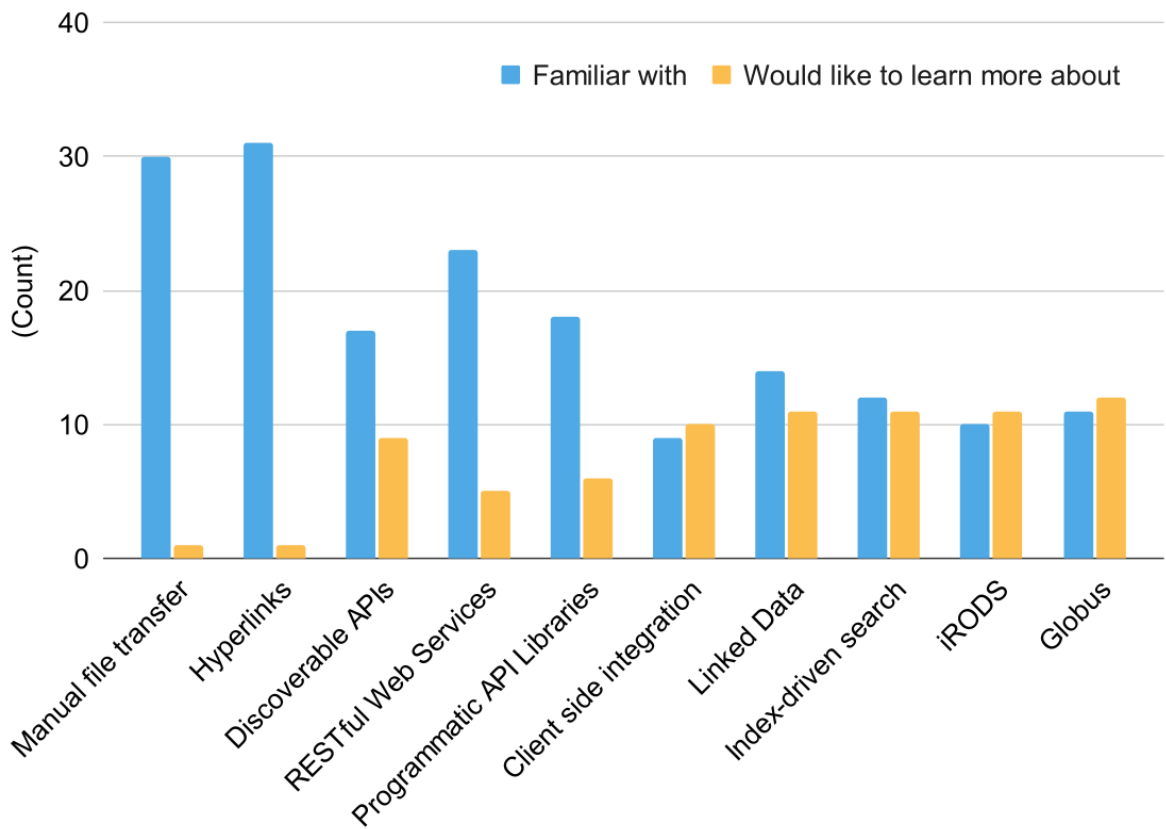

**Figure 7.** Familiarity with data sharing technologies (blue) and data sharing technology knowledge gaps (yellow). There were 32 responses to this question.

## Ontology Survey Results

Out of 32 responses to this question, 17 worked on databases with plant data, four on animals and 11 on databases with data from all organisms or other types of data (e.g., manure) (Fig. 8). Over 93% of the respondents report using ontologies or controlled vocabularies. In the majority of cases these ontologies are well-known and public such as the Gene Ontology (GO), Plant Ontology (PO), and the NCBI taxonomy, although over 30% of respondents reported using in-house ontologies or controlled vocabularies. In these cases, the ontology/controlled vocabulary was often species-specific such as SoyBase (SOY) or the Animal Trait Ontology (ATO), with efforts to associate terms with a more general ontology to facilitate cross-specific comparisons.

The ontologies are used to annotate both genomic (sequence, structure, and function) and phenomic data, as well as species (taxonomy) and breeds.

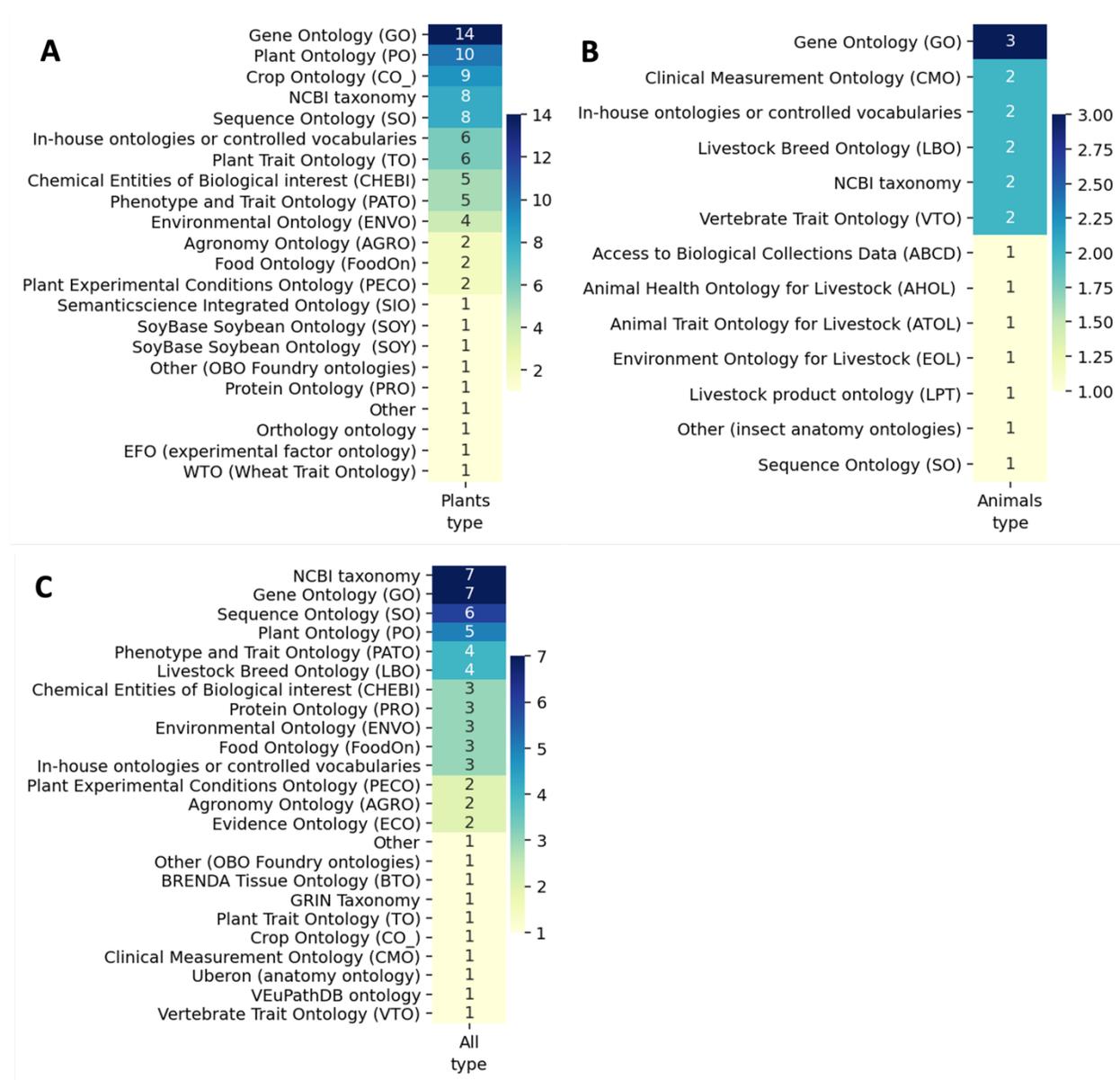

**Figure 8.** Use of Ontologies in AgBioData GGB Databases, showing the number of respondents that use a particular ontology in their database. Results are shown based on the focus of the database: (A) plant data; (B) animal data; or (C) databases storing both animal and plant data, as well as other types of data relevant to agricultural genetics, genomics and breeding.

## Tools used to explore, browse, contribute to, or develop ontologies or CVs

Of the 27 responses, the most commonly used sites to explore or browse ontologies were the EBI-OLS (48%; https://www.ebi.ac.uk/ols/index), followed by the Planteome (37%;

) and the OBO Foundry (30%; https://obofoundry.org/). A smaller proportion of respondents reported using Agroportal (26%; https://agroportal.lirmm.fr/) and Bioportal (19%; https://bioportal.bioontology.org/). Three respondents replied with none or not applicable.

When asked about tools for ontology development or maintenance, most respondents (14/29; 48%) report not using any. This may indicate that they are consumers of externally managed ontologies as opposed to creators of ontologies. This is supported by the high usage of externally managed ontologies reported by the respondents.

Of those who are editing or developing controlled vocabularies, the tools usage is concentrated on just three tools: OBO-Edit (33), Protège, and a web-enabled version of Protege, WebProtege (34). These programs all are open source and are free. OBO-Edit was designed to operate on ontologies using the OBO file format. The OBO file type is adequate to express simple relationships between terms such as "is_a", "part_of", "has_part", "regulates", etc. With its limited scope, OBO-Edit is easiest to learn and to apply to locally produced, species specific ontologies. Protège uses the more expressive file format, OWL. This program is designed to allow the user the ability to more formally express relationships between ontology terms. As a result, OWL files have the ability to be "reasoned" over and are thus more suitable for describing semantic relationships. This power and complexity comes with a hefty learning curve, making the format less suited for smaller ontology projects. The OWL format is currently utilized by a number of widely used ontologies, such as the GO, PO, Phenotype And Trait Ontology (PATO), Plant Trait Ontology (TO), etc.

About half of the respondents (55%, 18/33) are actively contributing to the ontologies they are using, with the primary method being via GitHub, closely followed by email. Other mechanisms reported were term request tools, web portals, and unspecified. Several respondents reported using a combination of these methods for term submission and feedback.

## Integration of ontologies/CVs into member databases

There were 31 responses to this question. Respondents reported using integrated ontologies for annotation, curation, searching, and archiving. The GO was the most often integrated (64.5%), followed by the NCBI Taxon Ontology (38.7%). Another twelve ontologies were integrated into more than one database: PO, Sequence Ontology (SO), Crop Ontology (CO), TO, PATO, Chemical Entities of Biological Interest (ChEBI), Livestock Breed Ontology (LBO), Clinical Measurement Ontology (CMO), Vertebrate Trait Ontology (VTO), Plant Experimental Conditions Ontology (PECO) and Evidence Ontology (ECO). Twenty-four other ontologies were each reported as being integrated by a single database. Two respondents reported that they don't integrate ontology information in their databases. One respondent said that they don't have any integrated ontologies currently but they are considering this for a new database in the future.

## Use of ontologies in search pages, or to pull/obtain the data from other data sources

There were 19 responses to this question. Two respondents answered 'NA' and one answered 'no'. Several ontologies were reported as being used in general text or keyword searches: GO (3), NCBI (3), PO (2), SO (2), LBO (2), VTO (2), CMO (2). The following ontologies were all mentioned once as being used to pull data from other data sources: BRENDA Tissue Ontology (BTO), CO, ECO, Experimental Factor Ontology (EFO), FlyBase Developmental Ontology (FBdv), Livestock Product Ontology (LPT), other insect ontologies, Wheat Trait Ontology (WTO), and Uberon.

Gene searches used GO (6), NCBI Taxonomy (1), PO (1) and CO (1). Marker/QTL/SNP searches used TO (2), VTO (1), LPT (1), CMO (1) and CO (1). Phenotype searches used PATO (2), ChEBI (2), PO (1), PECO (1), CO (1), GO (1), ENVO (1), NCBI (1) and TO (1). Genomics and microbiome data were searchable in one database using the VEuPathDB ontology.

One respondent addressed the second half of this question. They do not pull data from other sources, but they do provide links out to those sources. Exactly which sources were not specified in their response.

## Reasons for not using ontologies/CVs: Barriers and Needs

More than half of the respondents (nine out of 15 respondents) gave a reason for not using ontologies, and the majority of those were related to insufficient funding and expertise to get started. Only two respondents said that they didn't use ontologies because the ontologies themselves were either insufficient or too specific. When specifically asked if they had a need or gap that was not being addressed by ontologies, 60% of the respondents gave an answer, and the majority of those answers mentioned a gap in coverage for anatomy or phenotypes for specific organisms or groups of organisms. Only four respondents said that all of their needs were being met, and two were unsure.

Two of the more interesting answers were the inability to fully document mappings between ontologies and the inability to fully represent processes. These answers are interesting because there are newly developed methods for doing both. The SSSOM standard(35) records mappings with full provenance and the GO-CAM method allows for structured representation of biological processes (36). This suggests that information about new developments in semantic engineering are not being communicated to potential users.

When asked if they have experienced problems using ontologies for data sharing, the vast majority of respondents gave an answer and 63% said no; however, some of these answers are from people who are not using ontologies. About 46% of respondents had some sort of problem using ontologies for data sharing. These problems were about evenly split between issues with terms and issues with the costs of starting up. Specific issues with terms included the following:
- The presence of the same term in multiple ontologies causing confusion
- The time it takes to go from requesting a term to having a usable term is too long

- A required term does not exist or is hard to find
- It is hard to pick the right term

Problems with using ontologies for data sharing included those already mentioned, a lack of staff time and expertise, but also included the lack of easy-to-use tools and services for curating data, managing ontology versioning, and indexing for search. According to this survey, the major barriers to using ontologies include a lack of tools and services to support ontology use, lack of staff training, and terminology gaps in the ontologies themselves. The tools and services that do exist have become much easier to use over the past decade, but still require substantial training by staff to be usable. An investment in usability of ontologies and their tools and services, as well as a lower barrier to participation in the ontology development community (such as the OBO Academy (37)), could make a substantial impact.

## Training needs related to ontologies and shared vocabularies

There is a general need in the community for training in ontologies, specifically, best practices and matching of ontologies to needs (i.e., where/when to use which ontology). Out of 20 individuals responding to the question, six (30%) indicated they had no training needs. Of the training needs listed in the responses, the respondents mentioned the following: assessing whether an ontology is fit for the purpose; outreach on the uses and benefits of ontologies; general ontology development; tools development and assessment; and general training. Other notable training requests included training for developers as well as several specific areas of interest, ontology and annotation documentation, how to use an ontology in a database, technical training related to ontologies, and how to contribute to an ontology.

## Plans for using ontologies/ CVs in the future

The majority of respondents (86%) (out of 29 respondents) do have plans to use ontologies in the future, with an additional 6.9% stating that they may do so. Of those responding positively, the majority plan to use the GO and PATO (both 42.9%), the PO, CO, and PECO (all 23.8%), SO and ENVO (both 19%) and PRO (14%). The other ontologies and in house CVs represented small proportions.

# Summary of the results of the 2017 survey

The previous survey of AgBioData member databases in 2017 (31) focused more on the ontologies and less so on data sharing. The survey had a similar number of responses, 25 members representing about 29 different databases or resources.

The results showed the most common use of ontologies for curation was annotating sequence files with GO, SO, and PO and various trait ontologies (TO, VT, LPT) for QTLs, phenotypes and germplasm. The respondents mentioned 58 total ontologies of interest. Along with curation, respondents reported using ontologies (mostly GO) to search gene data and trait-related ontologies to search for QTLs, phenotypes, etc. from other sources.

Despite the existence of data standards for curation and software tools, the challenges in applying the ontologies for data curation also centered around the lack of funding for biocuration work at the member databases, lack of suitable tools, and a lack of specific publicly available ontologies for certain tasks, forcing data resources to develop in-house vocabularies to address these needs. Thus, it mirrors the present ontology survey results.

# CONCLUSIONS

The purpose of this survey was to assess the data sharing and ontology needs of the agriculture science community and the most impactful role AgBioData could play in increasing data sharing and ontology use across databases. Overall, the survey received responses from the majority of the AgBioData database community - 37 databases responded, and there were 44 databases or resources in the Consortium at the time. Our survey was targeted towards database personnel, and our goal was to receive at least one response per database. As such, we are fairly confident that the answers in the survey are largely representative of the community. Evaluation of the responses has allowed us to make several recommendations on topics the AgBioData Consortium could focus on in the future in order to improve member databases' data sharing and ontology use.

***Data sharing training for database personnel.*** Overall, survey respondents were aware of most data sharing techniques. However, the survey exposed several areas where training is desired: Discoverable APIs; Linked-Data; Client-side integration of results from multiple data sources; Index-driven search technologies; Data Management Systems; and Data Sharing via services (e.g., Globus). The AgBioData consortium should consider focusing on providing training resources in those areas for database personnel.

***Assessment of the extent of and barriers to metadata sharing.*** Survey respondents imply that there is a high level of data sharing among AgBioData databases - this result is comparable to a similar question from an AgBioData survey in 2017. However, the method of sharing, and the types of data that are shared, is heterogeneous in the current results. In particular, we recognized from the survey results that there may be barriers to metadata sharing. We suggest that future AgBioData working groups study how well metadata is shared among databases.

***Stakeholder education on the benefits of data sharing.*** Survey respondents thought that improved data sharing and discoverability was very important, similar to survey results from 2017. However, actually spending time and resources on improving them is likely less of a development priority due to lack of user awareness of its importance. We suggest that promoting an understanding of data sharing and discoverability in the user/stakeholder community should be a high future priority for AgBioData.

***Focus on improvements to phenotypic data sharing.*** Seven out of 37 databases stated that phenotypic and phenomic data are still challenging to share - whether providing the data from their own database or pulling from other databases. Phenotypic data represents an extremely diverse class of data - AgBioData should prioritize improvements for specific phenotypic data types and formats in future working groups.

**Lower specific barriers to data sharing.** Funding (62.5%), and time and resources (81.3%) were cited as the main barriers to data sharing. The AgBioData's Sustainability working group may help provide solutions to funding problems for databases, by communicating funding needs for AgBioData databases to policymakers. Technical knowledge (47%) and lack of data standards (50%) were two additional, major barriers that AgBioData could facilitate the improvement of. Previous survey questions revealed specific methodologies that AgBioData could provide training for. Identification, promotion, or development of data standards - similar to work that the AgBioData GFF3 working group has performed before (38) could also be a high priority.

**Ontology Training and support for biocurators:** It is evident from the survey responses that the same barriers exist now that did 5 years ago: a lack of expert biocurator resources, funding to support them, suitable tools to lower the learning activation barrier and an overall understanding of the importance of ontology use. There are several resources available now that can help users who are interested in learning more. Many users are familiar with GitHub as a repository of the ontologies, a change since the last survey. The OBO Foundry (https://obofoundry.org/) and the EBI-OLS (https://www.ebi.ac.uk/ols4) both offer standards, tools and lists of ontologies by domain. The Planteome has developed a system where the species-specific Crop Ontology vocabularies are mapped to the reference Trait Ontology in order to provide a unified overview, while still maintaining the breeder Trait Dictionaries. The OBO Foundry (https://obofoundry.org/resources) lists a number of tools, resources and tutorials for ontology users and developers.

# APPENDICIES

**Appendix 1.** Survey instructions, definitions, and questions, from the AgBioData Ontologies and Data Federation Working Groups Survey - 2022

# ACKNOWLEDGEMENTS

The AgBioData consortium is supported by the National Science Foundation [2126334]. The authors would like to acknowledge support from USDA NIFA (2021-77039-35992 JLC), USDA-ARS (Planteome team), and the National Science Foundation (2126334, 1940330, and 1340112 AET & LDC). This work was supported by The Arabidopsis Information Resource (TAIR) which is funded by academic, institutional, corporate, and individual subscriptions; TAIR is administered by the 501(c)(3) non-profit Phoenix Bioinformatics (TZB). Brandon Whitehead was supported by New Zealand's Ministry of Business Innovation and Employment (MBIE) Infrastructure Platform. We would like to thank the CGIAR Research Initiative on Digital Innovation. This work was supported in part by the U.S. Department of Agriculture, Agricultural Research Service. Mention of trade names or commercial products in this publication is solely for the purpose of providing specific information and does not imply recommendation or endorsement by the U.S. Department of Agriculture. USDA is an equal opportunity provider and employer. We would like to thank the CGIAR Research Initiative on Digital Innovation.

# Appendix 1: Survey Questions

## AgBioData Ontologies and Data Federation Working Groups Survey- 2022

This survey is designed to:
  1. Assess the current and future state of ontology use and data sharing utilized by the AgBioData community
  2. Collect data sharing use cases
Our goal is to identify areas that can be targeted for further development and have the most impact on the community.

YOUR PARTICIPATION IS VOLUNTARY AND APPRECIATED.

The survey should no longer than 15 minutes

## Definitions:

Note: If you are familiar with this material, skip to section 2.

Data sharing
Data Sharing is a generic term covering all methodologies and technologies for passing information from one system to another, to be used by another person, tool, or calculation. Data Sharing methodologies exist on a spectrum of automation. At one end of the spectrum there is very manual sharing like shipping files to a collaborator on a hard drive. On the other end of the spectrum there is full automation where data can travel between software systems freely and automatically, without human intervention.

Programmatic Access
Programmatic Access is the ability to read and interact with data using software. If a system or database has Programmatic Access, then someone should be able to write a new piece of code which can access the data in an automated way.

Data Federation
Data Federation is a software model where a collection of databases become highly interoperable, to the point where they appear to be parts of a single system to the outside world. An external tool should be able to query all the databases in the federation in the same way. An external tool should also be able to retrieve connected, interoperable data from multiple sources in the federation. An end user may not even care which source the data comes from, as long as it is available somewhere in the federation.

Interoperability
Interoperability is measure of how easy it is for two or more systems to interact and share data and has three facets: semantics, syntax, and protocol.

• Semantic Interoperability: The ability of computer systems to exchange data with unambiguous, shared meaning. For example, in different systems, 'plant height' may mean the distance from the soil surface to the tip of the inflorescence, to the top leaf, or to the top of the stem, excluding leave

• Syntactic interoperability: The structure of data and how it is organized. For example, the taxa information might be organized "genus species sub-taxa" or "sub-taxa species genus" in different systems, even though they contain the same information.

• Protocol Interoperability: The mechanism for how data transfers from one place to another. For example, one system only provides file downloads and another system only provides an API. These two systems could use the same semantics and syntax, but are still not interoperable.

Application Programming Interface (API)
An API is the part of a system or database which provides Programmatic Access to the data. There are many different types of API, but typically (in this context) we are referring to Web Service API's, which use the Internet to provide Programmatic Access to the outside world.

Annotations: The association between ontology terms and data objects
Data annotation is the process of labeling data objects with ontology terms so that computers can understand the data.

Functional Annotations:
Functional annotation is the process of attaching biological information, related to the function of the gene or section of the genome, to sequences of genes or proteins, using ontology terms from the Molecular Function branch of the Gene Ontology.

Structural Annotations:
Structural annotations are labels applied to physical regions of a genome that encode a genomic feature. Examples of such annotations are genes, mRNA, transcript, repeat sequences, etc.

Data Sharing Spectrum

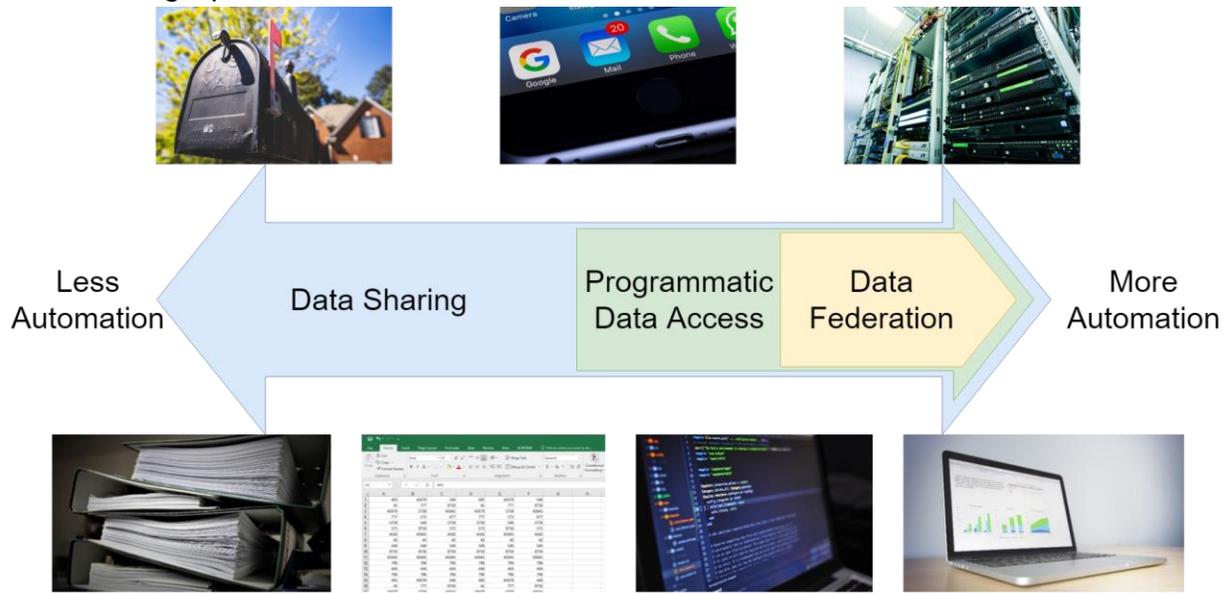

Ontology Spectrum

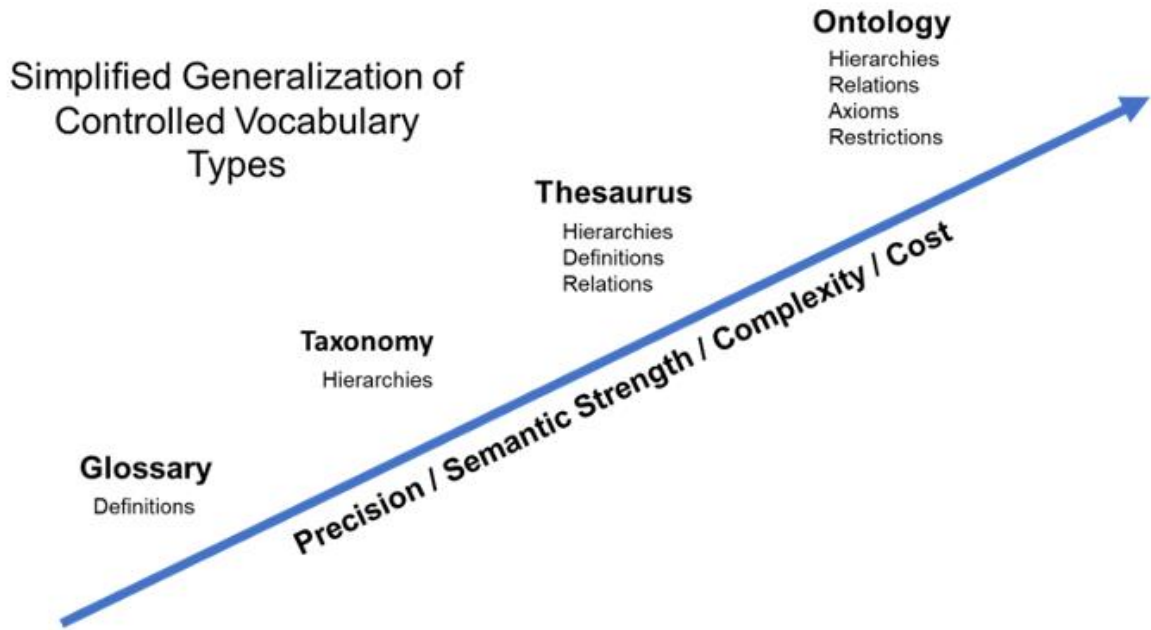

1.    1.1 Do you have any questions or comments about these definitions?
[Free text block]

# Survey participant information

2.    2.1 What database(s) or data resource(s) do you represent?
[Free text block]

3.    2.2 What are your roles in the online database/data resource? (choose all that apply)
Check all that apply.
Project PI
Project Manager
Computational Biologist
Developer
Curator
Maintainer
User
Other: [free text line]

4.    2.3 What types of data do you work with, generate, use, or store? (choose all that apply)
Check all that apply.
Reference genome sequence data
Structural and functional annotations
Transcriptome data (e.g., RNASeq)
DNA resequencing data (GBS, etc)
Genome Wide Association Studies (GWAS)
Gene, mRNA, protein (e.g., imported from public repositories or user submitted)
Molecular markers
QTL
Genetic maps
Germplasm/breeding lines
Phenotypes, and their correlation with specific genotypes
Genotypes (SNP's, alleles, etc) and genetic variation
Pangenomes
Comparative genomics
Other: [free text line]

5.    2.4 Do you use ontologies or controlled vocabularies in your work?
Mark only one oval.
Yes
No    Skip to question 14
I'm not sure

# Technologies and tools in use: Ontologies

6.    3.1 Please indicate which ontologies are currently used in your work or system:
Check all that apply.
Gene Ontology (GO)
Plant Ontology (PO)
Plant Trait Ontology (TO)
Plant Experimental Conditions Ontology (PECO)
Crop Ontology (CO_)
Phenotype and Trait Ontology (PATO)
Sequence Ontology (SO)
Environmental Ontology (ENVO)
Chemical Entities of Biological interest (CHEBI)
Protein Ontology (PRO)
Agronomy Ontology (AGRO)
Food Ontology (FoodOn)
NCBI taxonomy
Livestock Breed Ontology (LBO)
In-house ontologies or controlled vocabularies
Other: [free text line]

7.    3.2 Please specify which in-house ontologies or controlled vocabularies are
currently used. Add a link, if available.
[free text block]

8.    3.3 Please describe what types of data are annotated with the ontologies selected
above. For example "Plant Trait Ontology is used to annotate germplasm data and
phenotypic observations, the Gene Ontology is used with molecular marker data" etc
[free text block]

9.    3.4 What tools or sites do you use to explore or browse ontology terms?
Check all that apply.
Planteome web site or GitHub
EBI-OLS
OBO Foundry
BioPortal
AgroPortal
Other: [free text line]

10.    3.5 What ontology development, maintenance or exploration tools do you use, if any? (check all that apply)

Check all that apply.
OBO-Edit
Protégé
WebProtégé
NeOn Toolkit
SWOOP
Neologism
TopBraid Composer
Vitro
OWLGrEd
Other tools used: [free text line]
Not using any ontology tools

11.    3.6 Are you contributing to existing ontologies, such as requesting terms or providing feedback.? If so, how are you doing that (i.e. through GitHub, by email to contact person, etc)?
[free text block]

12.    3.7 Are ontologies integrated in your database or repository? Which ones and for what purposes?
[free text block]

13.    3.8 If you currently use ontologies in search pages in your database, please list the ontology and the type of search pages (e.g., Gene Ontology for gene search page). If you currently use ontologies to pull/obtain the data from other data sources, please list the ontology and data source.
[free text block]

If you are not currently using Ontologies:

14.    4.1 What are your major reasons for not using ontologies right now? (annotation, search pages, data retrieval from other sources)
[free text block]

15.    4.2 Do you have any training needs related to ontologies and shared vocabularies? If yes, please specify.
[free text block]

## Desired technology and tools: ontologies

16.    5.1 Do you plan to use ontologies in the future?
Mark only one oval.
Yes
No
Maybe

17.    5.2 Please indicate which ontologies you would like to begin using in your system
Check all that apply.
Gene Ontology (GO)
Plant Ontology (PO)
Plant Trait Ontology (TO)
Plant Experimental Conditions Ontology (PECO)
Crop Ontology (CO_)
Phenotype and Trait Ontology (PATO)
Sequence Ontology (SO)
Environmental Ontology (ENVO)
Chemical Entities of Biological interest (CHEBI)
Protein Ontology (PRO)
Agronomy Ontology (AGRO)
Food Ontology (FoodOn)
NCBI taxonomy
Livestock Breed Ontology (LBO)
In-house ontologies or controlled vocabularies
Other: [free text line]

18.    5.3 If you plan to use ontologies to pull/obtain the data from other data sources in
the future, please list the ontology and data source.
[free text block]

Data sharing in use

19.    6.1 Does your database share data with other databases, systems, or tools?
Mark only one oval.
Yes, data can be shared generically with any potential consumer
Yes, but it is designed to be shared only with specific tools
No, the capability for sharing is available but no other systems are consuming it yet
Skip to question 23
No, it would take significant effort to share the data       Skip to question 23

# Technologies and tools in use: data sharing

20.   7.1 What mechanism(s) do you currently use for sharing?
Check all that apply.
Manual transfer of flat files (FTP, email, Dropbox etc)
Hyperlinks to flat files
Discoverable Web Service APIs
Shared search indices
Direct SQL access (or alternate query language, e.g. GraphQL)
Other: [free text line]

21.   7.2 What data types do you share with other databases?
Check all that apply.
raw sequence files (e.g., FASTA)
variant files (e.g., SNP data, QTLs, VCF)
annotation/ontology files (e.g., GFF3)
genome read alignment files (e.g., SAM, BAM)
transcriptomic (e.g., MAGE-TAB)
phenomic (e.g., traits, images)
marker data (e.g., PCR, SSR)
Other: [free text line]

22.   7.3 What types of metadata information do you share with other databases
Check all that apply.
data-associated publication(s)
environmental
original data source
material and methods of data collection
data collection date
Other: [free text line]

Awareness of tools and technologies- data sharing

23.   8.1 What data sharing technologies are you familiar with? What data sharing
technologies would you appreciate learning more about?
Mark only one oval per row.

|  | Familiar with (i.e., working knowledge) | Would like to learn more about |
|---|---|---|
| Manual file import/export |  |  |
| Hyperlinks, persistent identifiers (e.g., DOIs) |  |  |
| Discoverable APIs (e.g., registries, service brokering) |  |  |
| RESTful Web Service APIs |  |  |
| Programmatic API Library (R, Python, Java, etc) |  |  |
| Linked-Data (Semantic Web) |  |  |
| Client side integration of results from data sources (e.g. sharing genome browser tracks) |  |  |
| Index-driven search technologies (e.g. ElasticSearch, Solr) |  |  |
| Data Management System (e.g. iRODS) |  |  |
| Data Sharing via services (e.g. Globus) |  |  |

24.   8.2 Are there any other technologies not listed above that you would like to learn more about?
[free text block]

## Desired technology and tools: data sharing

25.    9.1 How important is it to you to make your database more discoverable and available?
Mark only one oval.

      1        2        3        4        5
      Not important at all         Very important

26.    9.2 How important is it to your user community to make your database more discoverable and available?
Mark only one oval.

      1   2   3        4        5
      Not important at all         Very Important

27.    9.3 How high is it in your development priorities to make your database more discoverable and available, given the financial and time cost associated with it?
Mark only one oval.

      1   2   3        4        5
      Very Low                    Very High

28.    9.4 Please comment on how you want to make your data more available. What technologies or services would you like to use or provide to make your data more discoverable and available? If you are unsure but open to suggestions, please indicate that as well.
[free text block]

29.    9.5 What are the types of data or metadata that you would like to be able to share with external users, organizations, or tools?
[free text block]

30.    9.6 What are the types of data or metadata that you wish other organizations would share with you, or make easier to access?
[free text block]

# Barriers to success

31.　10.1 Do you have a need for ontologies that is not being met? Is there a gap in the available ontologies?

[free text block]

32.　10.2 Have you experienced problems using ontologies for data sharing?

Mark only one oval.

Yes

No

Sort of (explain below)

33.　Explain

34.　10.3 What do you feel are the biggest blockers to successful data sharing in your community?

*Check all that apply.*

Technical Knowledge

Lack of Data Standards

Data Licensing Issues

Funding / ROI

Time and Resources

Institutional Policy

Sharing Across Borders (international borders, institutional borders, etc)

Other: [free text line]

35.　10.4 If possible, please provide an example where you have experienced one of the blockers above?

[free text block]

36.　10.5 If you represent a database provider, do you or have you attempted to import, link, or share data programmatically from another database?

Mark only one oval.

Yes

No

37.　10.6 Were you successful? Why or why not?

[free text block]

38.  10.7 What types of data do you wish you could share more effectively from your database?
[free text block]

39.  10.8 What types of data do you wish you could access from other databases?
[free text block]

40.  10.9 - If you have detailed use cases regarding barriers to data sharing that you'd like to share with us and the community, feel free to post them in our discussion forum: https://github.com/AgBioData/DataFederation_WG/discussions
[free text line]

Additional thoughts and comments

41.  11.1 What areas of data sharing and/or ontologies do you wish this survey covered, but it didn't?
[free text block]

42.  11.2 Any other thoughts or comments?
[free text block]